\documentclass[12pt]{article}
\usepackage[english]{babel}
\usepackage{amsmath}
\usepackage{graphicx,psfrag,epsf}
\usepackage{enumerate}
\usepackage{natbib}
\usepackage{amsfonts}
\usepackage[latin1]{inputenc}
\usepackage[ulem=normalem]{changes}
\usepackage{cancel}
\usepackage{paralist}
\usepackage{multirow}
\usepackage[small]{caption}
\usepackage{setspace}
\usepackage{url}
\usepackage{bbm}
\usepackage{dsfont}
\usepackage{sectsty}
\usepackage{lipsum}
\usepackage{epstopdf}
\usepackage{mathtools}
\DeclarePairedDelimiter\ceil{\lceil}{\rceil}

\usepackage{amssymb}

\newcommand{\blind}{0}

\sectionfont{\centering}
\newtheorem{theorem}{Theorem}
\newtheorem{lemma}{Lemma}

\begin{document}

\def\spacingset#1{\renewcommand{\baselinestretch}%
{#1}\small\normalsize} \spacingset{1}

\if0\blind
{
  \title{\bf Functional Outlier Detection by a Local Depth with Application to NO$_{x}$ Levels}
  \author{
    \textbf{Carlo Sguera}\\
    \normalsize Department of Statistics, Universidad Carlos III de Madrid\\
    \normalsize 28903 Getafe (Madrid), Spain\\
    \normalsize(\texttt{csguera@est-econ.uc3m.es})\\
   \textbf{Pedro Galeano} \\
    \normalsize Department of Statistics, Universidad Carlos III de Madrid\\
    \normalsize 28903 Getafe (Madrid), Spain\\
    \normalsize(\texttt{pedro.galeano@uc3m.es})\\
     and \\
    \textbf{Rosa E. Lillo} \\
    \normalsize Department of Statistics, Universidad Carlos III de Madrid\\
    \normalsize 28903 Getafe (Madrid), Spain\\
    \normalsize(\texttt{rosaelvira.lillo@uc3m.es})\\
}
    \date{}
  \maketitle
} \fi

\if1\blind
{
  \bigskip
  \bigskip
  \bigskip
  \begin{center}
    {\LARGE\bf Functional Outlier Detection by a Local Depth with Application to NO$_{x}$ Levels}
\end{center}
  \medskip
} \fi

\begin{abstract}
This paper proposes methods to detect outliers in functional data sets and the task of identifying atypical curves is carried out using the recently proposed kernelized functional spatial depth (KFSD). KFSD is a local depth that can be used to order the curves of a sample from the most to the least central, and since outliers are usually among the least central curves, we present a probabilistic result which allows to select a threshold value for KFSD such that curves with depth values lower than the threshold are detected as outliers. Based on this result, we propose three new outlier detection procedures. The results of a simulation study show that our proposals generally outperform a battery of competitors. We apply our procedures to a real data set consisting in daily curves of emission levels of nitrogen oxides (NO$_{x}$) since it is of interest to identify abnormal NO$_{x}$ levels to take necessary environmental political actions.
\end{abstract}

\noindent%
\vfill
\hfill

\newpage
\spacingset{1.5}

\section{INTRODUCTION}
\label{sec:intro}
The accurate identification of outliers is an important aspect in any statistical data analysis. Nowadays there are well-established outlier detection techniques in the univariate and multivariate frameworks (for a complete review of the topic, see for example \citeauthor{BarLew1994} \citeyear{BarLew1994}). In recent years, new types of data have become available and tractable thanks to the evolution of computing resources, e.g., big multivariate data sets having more variables than observations (high-dimensional multivariate data) or samples composed of repeated measurements of the same observation taken over an ordered set of points that can be interpreted as realizations of stochastic processes (functional data). In this paper we focus on functional data, which are usually studied with the tools provided by functional data analysis (FDA). For overviews on FDA methods, see \cite{RamSil2005}, \cite{FerVie2006}, \cite{HorKok2012} or \cite{Cue2014}. For environmental statistical problems tackled using FDA techniques, see for example \cite{IgnEtAl2015}, \cite{MenEtAl2014} and \cite{RuiEsp2012}.

As in univariate or multivariate analysis, the detection of outliers is also fundamental in FDA. According to \citeauthor{FebGalGon2007}\ (\citeyear{FebGalGon2007}, \citeyear{FebGalGon2008}), a functional outlier is a curve generated by a stochastic process with a different distribution than the one of normal curves. This definition covers many types of outliers, e.g., magnitude outliers, shape outliers and partial outliers, i.e., curves having atypical behaviors only in some segments of the domain. Shape and partial outliers are typically harder to detect than magnitude outliers (in the case of high magnitude, outliers can even be recognized by simply looking at a graph), and therefore entail more challenging outlier detection problems. In this paper we focus on samples contaminated by low magnitude, shape or partial outliers.

Specifically, we consider a real data set consisting in nitrogen oxides (NO$_{x}$) emission daily levels measured in the Barcelona area (see \citeauthor{FebGalGon2008} \citeyear{FebGalGon2008} for a first analysis of this data set). Since NO$_{x}$ represent one of the most important pollutants, cause ozone formation and contribute to global warning, it is of interest the identification of days with abnormally large NO$_{x}$ emissions to allow the implementation of actions able to control their causes, which are primarily the combustion processes generated by motor vehicles and industries.

We propose to detect functional outliers using the notion of functional depth. A functional depth is a measure providing a $P$-based center-outward ordering criterion for observations of a functional space $\mathbb{H}$, where $P$ is a probability distribution on $\mathbb{H}$. When a sample of curves is available, a functional depth orders the curves from the most to the least central according to their depth values and, if any outlier is in the sample, its depth is expected to be among the lowest values. Therefore, it is reasonable to build outlier detection methods that use functional depths.

\begin{figure}[!htbp]
\centering
\includegraphics[scale=.4]{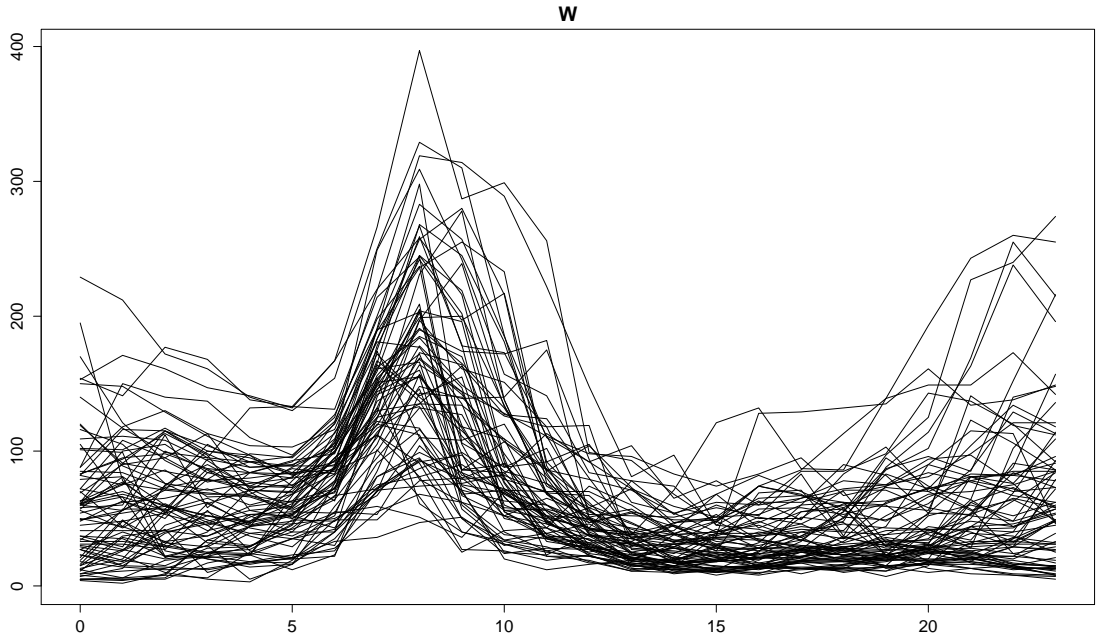}
\captionsetup{width=0.8\textwidth}
\caption{NO$_{x}$ levels measured in $\mu g/m^{3}$ every hour of 76 working days between 23/02/2005 and 26/06/2005 in Poblenou, Barcelona.}
\label{fig:NOxW}
\end{figure}

\indent In this paper we enlarge the number of available functional outlier detection procedures by presenting three new methods based on a specific depth, the kernelized functional spatial depth (KFSD, \citeauthor{SguGalLil2014}\ \citeyear{SguGalLil2014}). KFSD is a local-oriented depth, that is, a depth which orders curves looking at narrow neighborhoods and giving more weight to close than distant curves. Its approach is opposite to what global-oriented depths do. Indeed, any global depth makes depend the depth of a given curve on the whole rest of observations, with equal weights for all of them. This is the case of a global-oriented depth such as the functional spatial depth (FSD, \citeauthor{ChaCha2014} \citeyear{ChaCha2014}), of which KFSD is its local version. A local depth such as KFSD may result useful to analyze functional samples having a structure deviating from unimodality or symmetry. Moreover, the local approach behind KFSD proved to be a good strategy in supervised classification problems with groups of curves not extremely clear-cut (see \citeauthor{SguGalLil2014} \citeyear{SguGalLil2014}). Alternatively, we illustrate that KFSD ranks well low magnitude, shape or partial outliers, that is, their corresponding KFSD values are in general lower than those of normal curves. Then, we propose different procedures to select a threshold for KFSD to distinguish between normal curves and outliers. These procedures employ smoothing resampling techniques and are based on a theoretical result which allows to obtain a probabilistic upper bound on a desired false alarm probability of detecting normal curves as outliers. Note that the probabilistic foundations of the proposed methods represent a novelty in FDA outlier detection problems. We study the performances of our procedures in a simulation study and analyzing the NO$_{x}$ data set. We show this data set in Figure \ref{fig:NOxW}, where it is already possible to appreciate that the presence of partial outliers is an issue. 

We compare our methods with some alternative outlier detection procedures: \cite{FebGalGon2008} proposed to label as outliers those curves with depth values lower than a certain threshold. As functional depths, they considered the Fraiman and Muniz depth (\citeauthor{FraMun2001} \citeyear{FraMun2001}), the h-modal depth (\citeauthor{CueFebFra2006} \citeyear{CueFebFra2006}) and the integrated dual depth (\citeauthor{CueFra2009} \citeyear{CueFra2009}). To determine the depth threshold, they proposed two different bootstrap procedures based on depth-based trimmed or weighted resampling, respectively; \cite{SunGen2011} introduced the functional boxplot, which is constructed using the ranking of curves provided by the modified band depth (\citeauthor{LopRom2009} \citeyear{LopRom2009}). The proposed functional boxplot detects outliers using a rule that is similar to the one of the standard boxplot; \cite{HynSha2010} proposed to reduce the outlier detection problem from functional to multivariate data by means of functional principal component analysis (FPCA), and to use two alternative multivariate techniques on the scores to detect outliers, i.e., the bagplot and the high density region boxplot, respectively. 

The remainder of the article is organized as follows. In Section \ref{sec:kfsd} we recall the definition of KFSD. In Section \ref{sec:odp} we consider the functional outlier detection problem. In Theorem \ref{th:inPaper01} we present the result on which are based three new outlier detection methods which employ KFSD as depth function. In Section \ref{sec:simStudy} we report the results of our simulation study, whereas in Section \ref{sec:NOx} we perform outlier detection on the NO$_{x}$ data set. In Section \ref{sec:conc} we draw some conclusions. Finally, in the Appendix we report a sketch of the proof of Theorem \ref{th:inPaper01}.

\section{THE KERNELIZED FUNCTIONAL SPATIAL DEPTH}
\label{sec:kfsd}
In functional spaces a depth measure has the purpose of measuring the degree of centrality of curves relative to the distribution of a functional random variable. Various functional depths have been proposed following two alternative approaches: a global approach, which implies that the depth of an observation depends equally on all the observations allowed by $P$ on $\mathbb{H}$, and a local approach, which instead makes depend the depth of an observation more on close than distant observations. Among the existing global-oriented depths there is the Fraiman and Muniz depth (FMD, \citeauthor{FraMun2001} \citeyear{FraMun2001}), the random Tukey depth (RTD, \citeauthor{CueNie2008} \citeyear{CueNie2008}), the integrated dual depth (IDD, \citeauthor{CueFra2009} \citeyear{CueFra2009}), the modified band depth (MBD, \citeauthor{LopRom2009} \citeyear{LopRom2009}) or the functional spatial depth (FSD, \citeauthor{ChaCha2014} \citeyear{ChaCha2014}). Proposals of local-oriented depths are instead the h-modal depth (HMD, \citeauthor{CueFebFra2006}\ \citeyear{CueFebFra2006}) or the kernelized functional spatial depth (KFSD, \citeauthor{SguGalLil2014}\ \citeyear{SguGalLil2014}).

In this paper we focus on KFSD. Before giving its definition, we recall the definition of the functional spatial depth (FSD, \citeauthor{ChaCha2014} \citeyear{ChaCha2014}). Let $\mathbb{H}$ be an infinite-dimensional Hilbert space, then for $x \in \mathbb{H}$ and the functional random variable $Y \in \mathbb{H}$, FSD of $x$ relative to $Y$ is given by

\begin{equation*}
FSD(x,Y) = 1 - \left\|\mathbb{E}\left[\frac{x-Y}{\left\|x-Y\right\|}\right]\right\|,
\end{equation*}

\noindent where $\|\cdot\|$ is the norm inherited from the usual inner product in $\mathbb{H}$. For a $n$-size random sample of $Y$, i.e., $Y_{n} = \left\{y_{1}, \ldots, y_{n}\right\}$, the sample version of FSD has the following form:

\begin{equation}
\label{eq:sampleFSD}
FSD(x,Y_{n}) = 1 - \frac{1}{n}\left\|\sum_{i=1}^{n}\frac{x-y_{i}}{\left\|x-y_{i}\right\|}\right\|.
\end{equation}

As mentioned before, FSD is a global-oriented depth and KFSD is a local version of it. KFSD is obtained writing \eqref{eq:sampleFSD} in terms of inner products and then replacing the inner product function with a positive definite and stationary kernel function. This replacement exploits the relationship 

\begin{equation}\label{eq:kappa_phi}
\kappa(x,y) = \langle\phi(x), \phi(y)\rangle, \quad x, y \in \mathbb{H}, 
\end{equation}

\noindent where $\kappa$ is the kernel $\kappa: \mathbb{H} \times \mathbb{H} \rightarrow \mathbb{R}$, $\phi$ is the embedding map $\phi: \mathbb{H} \rightarrow \mathbb{F}$ and $\mathbb{F}$ is a feature space. Indeed, a definition  of KFSD in terms of $\phi$ can be given, that is,

\begin{equation}
\label{eq:popKFSD}
KFSD(x,Y) = 1 - \left\|\mathbb{E}\left[\frac{\phi(x)-\phi(Y)}{\left\|\phi(x)-\phi(Y)\right\|}\right]\right\|,
\end{equation}

\noindent and it can be interpreted as a recoded version of $FSD(x,Y)$ since $KFSD(x,Y)=$ \linebreak $FSD(\phi(x),\phi(Y))$. The sample version of \eqref{eq:popKFSD} is given by

\begin{equation*}
KFSD(x,Y_{n}) = 1 - \frac{1}{n}\left\|\sum_{i=1}^{n}\frac{\phi(x)-\phi(y_{i})}{\left\|\phi(x)-\phi(y_{i})\right\|}\right\|.
\end{equation*}

\noindent Then, standard calculations (see Appendix) and \eqref{eq:kappa_phi} allow to provide an alternative expression of $KFSD(x,Y_{n})$, in this case in terms of $\kappa$:

\begin{gather}
KFSD(x, Y_{n}) = 1 - \nonumber \\
\label{eq:sampleKFSD}
\frac{1}{n} \left(\sum_{\substack{i,j=1; \\ y_{i} \neq x; y_{j} \neq x}}^{n}\frac{\kappa(x,x)+\kappa(y_{i},y_{j})-\kappa(x,y_{i})-\kappa(x,y_{j})}{\sqrt{\kappa(x,x)+\kappa(y_{i},y_{i})-2\kappa(x,y_{i})}\sqrt{\kappa(x,x)+\kappa(y_{j},y_{j})-2\kappa(x,y_{j})}}\right)^{1/2},
\end{gather}

\noindent Note that \eqref{eq:sampleKFSD} only requires the choice of $\kappa$, and not of $\phi$, which can be left implicit. As $\kappa$ we use the Gaussian kernel function given by

\begin{equation}
\label{eq:ker}
\kappa(x,y) = \exp\left(-\frac{\|x-y\|^2}{\sigma^2}\right),
\end{equation}

\noindent where $x, y \in \mathbb{H}$. In turn, \eqref{eq:ker} depends on the norm function inherited by the functional Hilbert space where data are assumed to lie, and on the bandwidth $\sigma$. Regarding $\sigma$, we initially consider 9 different $\sigma$, each one equal to 9 different percentiles of the empirical distribution of $\left\{\|y_{i}-y_{j}\|, y_{i}, y_{j} \in Y_{n}\right\}$. The first percentile is 10\%, and by increments of 10 we obtain the ninth percentile, i.e., 90\%. Note that the lower $\sigma$, the more local the approach, and therefore the percentiles that we use cover different degrees of KFSD-based local approaches:  strongly (e.g., 20\%), moderately (e.g., 50\%) and weakly (e.g., 80\%) local approaches. In Section \ref{sec:simStudy} we present a method to select $\sigma$ in outlier detection problems.

\begin{figure}[!htbp]
\centering
\includegraphics[scale=.4]{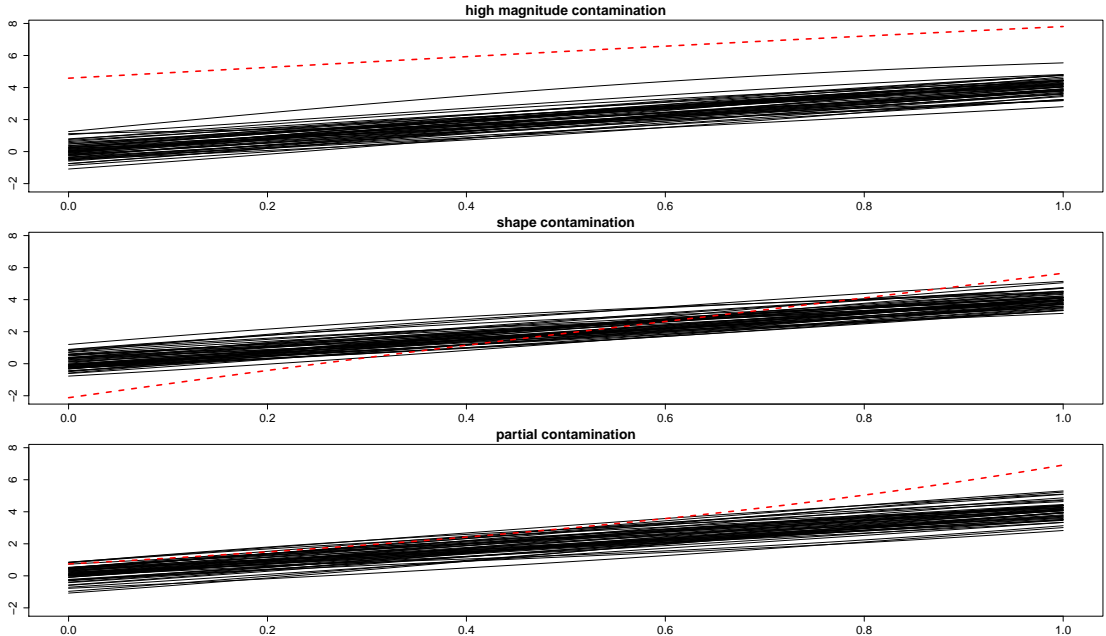}
\captionsetup{width=0.8\textwidth}
\caption{Examples of contaminated data sets: high magnitude contamination (top), shape contamination (middle) and partial contamination (bottom). The solid curves are normal curves and the dashed curves are outliers}
\label{fig:globVSloc}
\end{figure}

In general, since any functional depth measures the degree of centrality or extremality of a given curve relative to a distribution or a sample, outliers are expected to have low depth values. More in particular, in presence of low magnitude, shape or partial outliers, an approach based on the use of a local depth like KFSD may help in detecting outliers. To illustrate this fact, we present the following example: first, we generated 100 data sets of size 50 from a mixture of two stochastic processes, one for normal curves and one for high magnitude outliers, with the probability that a curve is an outlier equal to 0.05. Second, we generated a group of 100 data sets from a mixture which produces shape outliers. Finally, we generated a group of 100 data sets from a mixture which produces partial outliers. In Figure \ref{fig:globVSloc} we report a contaminated data set for each mixture. 

Let $n_{out,j}, j=1, \ldots, 100$, be the number of outliers generated in the $j$th data set. For each data set and functional depth, it is desirable to assign the $n_{out,j}$ lowest depth values to the $n_{out,j}$ generated outliers. For each mixture and generated data set, we recorded how many times the depth of an outlier is among the $n_{out,j}$ lowest values. As depth functions, we considered five global depths (FMD, RTD, IDD, MBD and FSD) and two local depths (HMD and KFSD). The results reported in Table \ref{tab:example} show that for all the functional depths the ranking of high magnitude outliers is an easier task than the ranking of shape and partial outliers. However, while the ranking of high magnitude outliers is reasonably good in different cases, e.g., for the local KFSD (94.87\%) and the global RTD (90.17\%), the ranking of shape and partial outliers is markedly better with local depths (shape: 86.72\% for KFSD and 85.47\% for HMD; partial: 82.03\% for KFSD and 81.25\% for HMD) than with the best global depths (shape: FSD with 39.06\%; partial: FSD with 46.48\%). These results suggest that, selecting a proper threshold, KFSD can isolate well outliers.

\begin{table}[!htbp]
\captionsetup{justification=justified,width=0.5\textwidth}
\caption{Percentages of times a depth assigns a value among the $n_{out,j}$ lowest ones to an outlier. Types of outliers: high magnitude, shape and partial.}
\centering
\scalebox{.75}{
\begin{tabular}{l|ccccc|cc}
\hline
type of depths & \multicolumn{5}{c}{global depths} \vline & \multicolumn{2}{c}{local depths} \\
\hline
depths & FMD & RTD & IDD & MBD & FSD & HMD & KFSD \\
\hline
high magnitude outliers & 86.32 & 90.17 & 81.62 & 69.23 & 68.80 & 85.47 & 94.87 \\ 
shape outliers & 7.81 & 33.59 & 38.67 & 12.11 & 39.06 & 85.94 & 86.72 \\  
partial outliers & 18.75 & 44.53 & 34.77 & 19.14 & 46.48 & 81.25 & 82.03\\
\hline
\end{tabular}}
\label{tab:example}
\end{table}

\section{OUTLIER DETECTION FOR FUNCTIONAL DATA}
\label{sec:odp}
The outlier detection problem can be described as follows: let $Y_{n}= \left\{y_{1}, \ldots, y_{n}\right\}$ be a sample generated from a mixture of two functional random variables in $\mathbb{H}$, one for normal curves and one for outliers, say $Y_{nor}$ and $Y_{out}$, respectively. Let $Y_{mix}$ be a mixture, i.e.,

\begin{equation}
\label{eq:mix}
Y_{mix} =
\left\{
	\begin{array}{ll}
		Y_{nor}, & \mbox{with probability } 1-\alpha, \\
		Y_{out}, & \mbox{with probability } \alpha,
	\end{array}
\right.
\end{equation}

\noindent where $\alpha \in [0,1]$ is the contamination probability  (usually, a value rather close to 0). The curves composing $Y_{n}$ are all unlabeled, and the goal of the analysis is to decide whether each curve is a normal curve or an outlier.\\ 
\indent KFSD is a functional extension of the kernelized spatial depth for multivariate data (KSD) proposed by \cite{CheDanPenBar2009}, who also proposed a KSD-based outlier detector that we generalize to KFSD: for a given data set $Y_{n}$ generated from $Y_{mix}$ and $t \in [0,1]$, the KFSD-based outlier detector for $x \in \mathbb{H}$ is given by

\begin{equation}
\label{eq:g_KFSD_no_b}
g(x,Y_{n}) =
\left\{
	\begin{array}{ll}
		1,  & \mbox{if } KFSD(x,Y_{n}) \leq t, \\
		0, 	& \mbox{if } KFSD(x,Y_{n}) > t,
	\end{array}
\right.
\end{equation}

\noindent where $t$ is a threshold which allows to discriminate between outliers (i.e., $g(x,Y_{n})=1$) and normal curves (i.e., $g(x,Y_{n})=0$), and it is a parameter that needs to be set.

For the multivariate case, KSD-based outlier detection is carried under different scenarios. One of them consists in an outlier detection problem where two samples are available and the threshold $t$ is selected by controlling the probability that normal observations are classified as outliers, i.e., the false alarm probability (FAP). The selection criterion is based on a result providing a KSD-based probabilistic upper bound on the FAP which depends on $t$. Then, the threshold for KSD is provided by the maximum value of $t$ such that the upper bound does not exceed a given desired FAP. We extend this result to KFSD:

\begin{theorem}\label{th:inPaper01}
Let $Y_{n_{Y}} = \left\{y_{i}, \ldots, y_{n_{Y}}\right\}$ and $Z_{n_{Z}} = \left\{z_{i}, \ldots, z_{n_{Z}}\right\}$ be i.\ i.\ d.\ samples generated from the unknown mixture of random variables $Y_{mix} \in \mathbb{H}$ described by (\ref{eq:mix}), with $\alpha > 0$. Let $g(\cdot,Y_{n_{Y}})$ be the outlier detector defined in (\ref{eq:g_KFSD_no_b}). Fix $\delta \in (0,1)$ and suppose that $\alpha \leq r$ for some $r \in [0,1]$. For a new random element $x$ generated from $Y_{nor}$, the following inequality holds with probability at least $1-\delta$:

\begin{equation}
\label{eq:ineThe01}
\mathbb{E}_{x | Y_{n_{Y}}}\left[g(x,Y_{n_{Y}})\right] \leq \frac{1}{1-r} \left[\frac{1}{n_{Z}} \sum_{i=1}^{n_{Z}} g\left(z_{i},Y_{n_{Y}}\right) + \sqrt{\frac{\ln{1/\delta}}{2 n_{Z}}} \right],
\end{equation} 

\noindent where $\mathbb{E}_{x | Y_{n_{Y}}}$ refers to the expected value of $x$ for a given $Y_{n_{Y}}$.
\end{theorem}

The proof of Theorem \ref{th:inPaper01} is presented in the Appendix. Recall that the FAP is the probability that a normal observation $x$ is classified as outlier. For the elements of Theorem \ref{th:inPaper01}, $\mathrm{Pr}_{x | Y_{n_{Y}}}\left(g(x,Y_{n_{Y}})=1\right)$ is the FAP. Moreover,

\begin{equation*}
\mathrm{Pr}_{x | Y_{n_{Y}}}\left(g(x,Y_{n_{Y}})=1\right) = \mathbb{E}_{x | Y_{n_{Y}}}\left[g(x,Y_{n_{Y}})\right]. 
\end{equation*}

\noindent Therefore, the probabilistic upper bound of Theorem \ref{th:inPaper01} applies also to the FAP.

It is worth noting that the application of Theorem \ref{th:inPaper01} requires to observe two samples, circumstance rather uncommon in classical outlier detection problems, in which usually a single sample generated from an unknown mixture of random variables is available. For this reason, we propose a solution which allows to use Theorem \ref{th:inPaper01} in presence of a unique sample. Note that the general idea behind holds also in the multivariate framework, and therefore it would enable to perform KSD-based outlier detection when only a $\mathbb{R}^{d}$-sample is available.

In the functional context, our solution consists in setting $Y_{n_{Y}} = Y_{n}$ and in obtaining $Z_{n_{Z}}$ by resampling with replacement from $Y_{n}$. Note that by doing this, and for sufficiently large values of $n_{Z}$, we also obtain that the effect of $\delta$ on the probabilistic upper bound drastically reduces. Concerning $r$, that is the upper bound for the unknown contamination probability $\alpha$, a true range between 0 and 0.1 appears to be appropriate to cover most of the situations found in practice. Regarding the resampling procedure to obtain $Z_{n_{Z}}$, we consider three different schemes, all of them with replacement. Since we deal with potentially contaminated data sets, besides simple resampling, we also consider two robust KFSD-based resampling procedures inspired by the work of \cite{FebGalGon2008}. The three resampling schemes that we consider are: 

\begin{compactenum}
\item\label{it:sim} Simple resampling.

\item\label{it:tri} KFSD-based trimmed resampling: once $KFSD(y_{i},Y_{n}), i=1,\ldots,n$ are obtained, it is possible to identify the $\ceil*{\alpha_{T}}\%$ least deepest curves, for a certain $0 < \alpha_{T} < 1$ usually close to 0, that we advise to set equal to $r$. These least deep curves are deleted from the sample, and simple resampling is carried out with the remaining curves.

\item\label{it:wei} KFSD-based weighted resampling: once $KFSD(y_{i},Y_{n}), i=1,\ldots,n$ are obtained, weighted resampling is carried out with weights $w_{i} = KFSD(y_{i},Y_{n})$.
\end{compactenum}

\noindent All the above procedures generate samples with some repeated curves. However, in a preliminary stage of our study we observed that it is preferable to work with $Z_{n_{Z}}$ composed of non-repeated curves. To obtain such samples, we add a common smoothing step to the previous three resampling schemes.

To describe the smoothing step, first recall that each curve in $Y_{n}$ is in practice observed at a discretized and finite set of domain points, and that the sets may differ from one curve to another. For this reason, the estimation of $Y_{n}$ at a common set of $m$ equidistant domain points may be required. Let $\left(y_{i}(s_{1}), \ldots,y_{i}(s_{m})\right)$ be the observed or estimated $m$-dimensional equidistant discretized version of $y_{i}$, $\Sigma_{Y_{n}}$ be the covariance matrix of the discretized form of $Y_{n}$ and $\gamma$ be a smoothing parameter. Consider a zero-mean Gaussian process whose discretized form has $\gamma\Sigma_{Y_{n}}$ as covariance matrix. Let $\left(\zeta(s_{1}), \ldots, \zeta(s_{m})\right)$ be a discretized realization of the previous Gaussian process. Consider any of the previous three resampling procedures and assume that at the $j$th trial, $j=1, \ldots, n_{Z}$, the $i$th curve in $Y_{n}$ has been sampled. Then, the discretized form of the $j$th curve in $Z_{n_{Z}}$ would be given by $\left(z_{j}(s_{1}), \ldots, z_{j}(s_{m})\right) = \left(y_{i}(s_{1})+\zeta(s_{1}), \ldots, y_{i}(s_{m})+\zeta(s_{m})\right)$, or, in functional form, by $z_{j} = y_{i} + \zeta$. Therefore, combining each resampling scheme with this smoothing step, we provide three different approximate ways to obtain $Z_{n_{Z}}$, and we refer to them as $smo$, $tri$ and $wei$, respectively. Then, for fixed $\delta$, $r$ and desired FAP, the threshold $t$ for \eqref{eq:g_KFSD_no_b} is selected as the maximum value of $t$ such that the right-hand side of \eqref{eq:ineThe01} does not exceed the desired FAP. Let $t^{*}$ be the selected threshold, which is then used in \eqref{eq:g_KFSD_no_b} to compute $g\left(y_{i}, Y_{n}\right)$, $i=1, \ldots, n$. If $g\left(y_{i}, Y_{n}\right) = 1$, $y_{i}$ is detected as outlier. To summarize, we provide three KFSD-based outlier detection procedures and we refer to them as KFSD$_{smo}$, KFSD$_{tri}$ and KFSD$_{wei}$ depending on how $Z_{n_{Z}}$ is obtained ($smo$, $tri$ and $wei$, respectively; recall that $Y_{n_{Y}}=Y_{n}$). As competitors of the proposed procedures, we consider the methods mentioned in Section \ref{sec:intro} that we now describe.\\
\indent \cite{SunGen2011} proposed a depth-based functional boxplot and an associated outlier detection rule based on the ranking of the sample curves that MBD provides. The ranking is used to define a sample central region, that is, the smallest band containing at least half of the deepest curves. The non-outlying region is defined inflating the central region by 1.5 times. Curves that do not belong completely to the non-outlying region are detected as outliers. The original functional boxplot is based on the use of MBD as depth, but clearly any functional depth can be used. Another contribution of this paper is the study of the performances of the outlier detection rule associated to the functional boxplot (from now on, FBP) when used together with the battery of functional depths mentioned in Section \ref{sec:kfsd}.\\
\indent \cite{FebGalGon2008} proposed two depth-based outlier detection procedures that select a threshold for FMD, HMD or IDD by means of two alternative robust smoothed bootstrap procedures whose single bootstrap samples are obtained using the above described $tri$ and $wei$, respectively. At each bootstrap sample, the 1\% percentile of empirical distribution of the depth values is obtained, say $p_{0.01}$. If $B$ is the number of bootstrap samples, $B$ values of $p_{0.01}$ are obtained. Each method selects as cutoff $c$ the median of the collection of $p_{0.01}$ and, using $c$ as threshold, a first outlier detection is performed. If some curves are detected as outliers, they are deleted from the sample, and the procedure is repeated until no more outliers are found (note that $c$ is computed only in the first iteration). We refer to these methods as B$_{tri}$ and B$_{wei}$, and also in this case we evaluate these procedures using all the functional depths mentioned in Section \ref{sec:kfsd}.\\
\indent Finally, we also consider two procedures proposed by \cite{HynSha2010} that are not based on the use of a functional depth. Both are based on the first two robust functional principal components scores and on two different graphical representations of them. The first proposal is the outlier detection rule associated to the functional bagplot (from now on, FBG), which works as follows: obtain the bivariate robust scores and order them using the multivariate halfspace depth (\citeauthor{Tuk1975} \citeyear{Tuk1975}). Define an inner region by considering the smallest region containing at least the 50\% of the deepest  scores, and obtain a non-outlying region by inflating the inner region by 2.58 times. FBG detects as outliers those curves whose scores are outside the non-outlying region. Note that the scores-based regions and outliers allow to draw a bivariate bagplot, which produces a functional bagplot once it is mapped onto the original functional space. The second proposal is related to a different graphical tool, the high density region boxplot (from now on, we refer to its associated outlier detection rule as FHD). In this case, once obtained the scores, perform a bivariate kernel density estimation. Define the $(1-\beta)$-high density region (HDR), $\beta \in (0,1)$, as the region of scores with coverage probability equal to $(1-\beta)$. FHD detects as outliers those curves whose scores are outside the $(1-\beta)$-HDR. In this case, it is possible to draw a bivariate HDR boxplot which can be mapped onto a functional version, thus providing the functional HDR boxplot.

\section{SIMULATION STUDY}
\label{sec:simStudy}
After introducing KFSD$_{smo}$, KFSD$_{tri}$ and KFSD$_{wei}$, their competitors (FBP, B$_{tri}$, B$_{wei}$, FBG and FHD), as well as seven different functional depths (FMD, HMD, RTD, IDD, MBD, FSD and KFSD), in this section we carry out a simulation study to evaluate the performances of the different methods. For FBP, B$_{tri}$ and B$_{wei}$, we use the notation procedure+depth: for example, FBP+FMD refers to the method obtained by using FBP together with FMD.\\
\indent To perform our simulation study, we consider six models: all of them generate curves according to the mixture of random variables $Y_{mix}$ described by \eqref{eq:mix}. The first three mixture models (MM1, MM2 and MM3) share $Y_{nor}$, with curves generated by 

\begin{equation}
\label{eq:Y_nor_123}
y(s) = 4s + \epsilon(s),
\end{equation}

\noindent where $s \in [0,1]$ and $\epsilon(s)$ is a zero-mean Gaussian component with covariance function given by

\begin{equation*}
\mathbb{E}(\epsilon(s),\epsilon(s^{\prime})) = 0.25 \exp{(-(s-s^{\prime})^2)}, \quad s,s^{\prime} \in [0, 1].
\end{equation*}

\begin{figure}[!htbp]
\centering
\includegraphics[scale=.4]{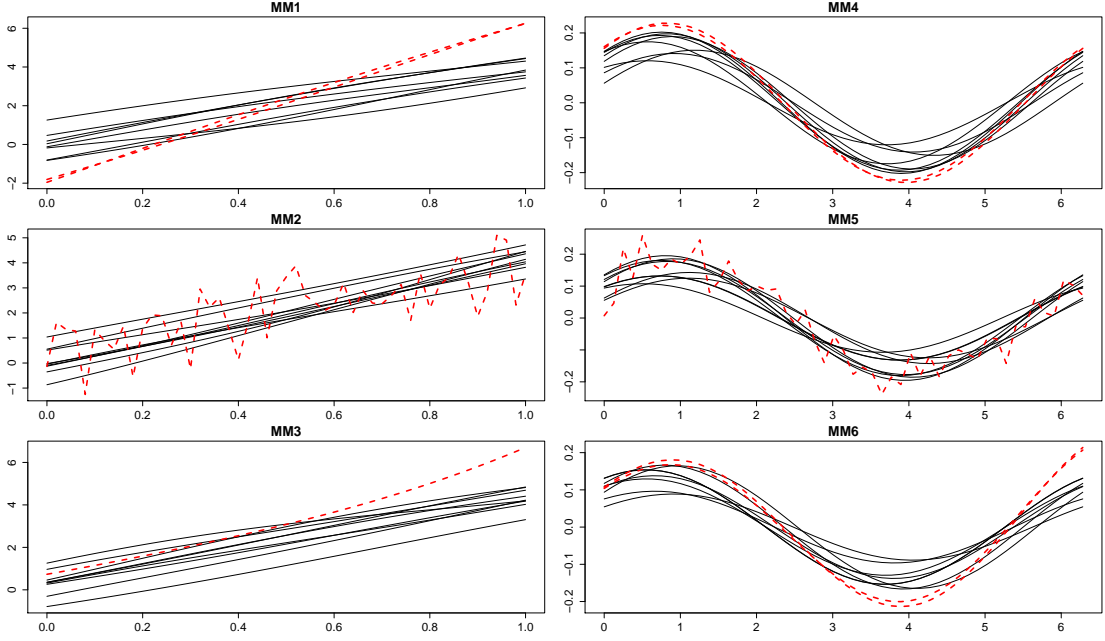}
\captionsetup{width=0.8\textwidth}
\caption{Examples of contaminated functional data sets generated by MM1, MM2, MM3, MM4, MM5 and MM6. Solid curves are normal curves and dashed curves are outliers.}
\label{fig:simStudy}
\end{figure}

\noindent Also the remaining three mixture models (MM4, MM5 and MM6) share $Y_{nor}$, but, in this case, the curves are generated by

\begin{equation}
\label{eq:Y_nor_456}
y(s) = u_{1}\sin s + u_{2}\cos s,
\end{equation}

\noindent where $s \in [0,2\pi]$ and $u_{1}$ and $u_{2}$ are observations from a continuous uniform random variable between 0.05 and 0.15.\\
\indent MM1, MM2 and MM3 differ in their $Y_{out}$ components. Under MM1, the outliers are generated by

\begin{equation*}
y(s) = 8s - 2 + \epsilon(s),
\end{equation*}

\noindent which produces outliers of both shape and low magnitude nature. Under MM2, the outliers are generated by adding to \eqref{eq:Y_nor_123} an observation from a $N(0,1)$, and as result outliers are more irregular than normal curves. Finally, under MM3, the outliers are generated by 

\begin{equation*}
y(s) = 4\exp(s) + \epsilon(s),
\end{equation*}

\noindent which produces curves that are normal in the first part of the domain, but that become exponentially outlying.\\
\indent Similarly, MM4, MM5 and MM6 differ in their $Y_{out}$ components. Under MM4, the outliers are generated replacing $u_{2}$ with $u_{3}$ in \eqref{eq:Y_nor_456}, where $u_{3}$ is an observation from a continuous uniform random variable between 0.15 and 0.17. This change produces partial low magnitude outliers in the first and middle part of the domain of the curves. Under MM5, the outliers are generated by adding to \eqref{eq:Y_nor_456} an observation from a $N(0,\left(\frac{0.1}{2}\right)^{2})$, and they turn out to be more irregular curves. Finally, under MM6, the outliers are generated by

\begin{equation}
\label{eq:Y_out_6}
y(s) = u_{1}\sin s + \exp\left(\frac{0.69s}{2\pi}\right) u_{4} \cos s,
\end{equation}

\noindent where $u_{4}$ is an observation from a continuous uniform random variable between 0.1 and 0.15. As MM3, MM6 allows outliers that are normal in the first part of the domain and become outlying with an exponential pattern. In Figure \ref{fig:simStudy} we report a simulated data set with at least one outlier for each mixture model.

The details of the simulation study are the following: for each mixture model, we generated 100 data sets, each one composed of 50 curves. As mentioned above, for each single samples Theorem \ref{th:inPaper01} cannot be directly applied, and therefore KFSD$_{smo}$, KFSD$_{tri}$ and KFSD$_{wei}$ represent practical alternatives. Two values of the contamination probability $\alpha$ were considered: 0.02 and 0.05. All curves were generated using a discretized and finite set of 51 equidistant points in the domain of each mixture model ($[0,1]$ for MM1, MM2 and MM3; $[0, 2\pi]$ for MM4, MM5 and MM6) and the discretized versions of the functional depths were used.\\
\indent In relation with the methods and the functional depths that we consider in the study, their specifications are described next:

\begin{compactenum}
\item FBP when used with FMD, HMD, RTD, IDD, MBD, FSD and KFSD: regarding FBP, as reported in Section \ref{sec:odp}, the central region is built considering the 50\% deepest curves and the non-outlying region by inflating by 1.5 times the central region. Regarding the depths, for HMD, we follow the recommendations in \cite{FebGalGon2008}, that is, $\mathbb{H}$ is the $L^2$ space, $\kappa(x,y) = \frac{2}{\sqrt{2\pi}}\exp\left(-\frac{\|x-y\|^2}{2h^2}\right)$ and $h$ is equal to the 15\% percentile of the empirical distribution of $\left\{\|y_{i}-y_{j}\|, y_{i}, y_{j} \in Y_{n}\right\}$. For RTD and IDD, we work with 50 projections in random Gaussian directions. For MBD, we consider bands defined by two curves. For FSD and KFSD, we assume that the curves lie in the $L^{2}$ space. Moreover, in KFSD we set $\sigma$ equal to a moderately local percentile (50\%) of the empirical distribution of $\left\{\|y_{i}-y_{j}\|, y_{i}, y_{j} \in Y_{n}\right\}$.

\item B$_{tri}$ and B$_{wei}$ when used with FMD, HMD, RTD, IDD, MBD, FSD and KFSD: $\gamma=0.05$, $B=200$, $\alpha_{T}=\alpha$. Regarding the depths, we use the specifications reported for FBP.

\item FBG: as reported in Section  \ref{sec:odp}, the central region is built considering the 50\% deepest bivariate robust functional principal component scores and the non-outlying region by inflating by 2.58 times the central region.

\item FHD: $\beta = \alpha$.

\item KFSD$_{smo}$, KFSD$_{tri}$ and KFSD$_{wei}$: $n_{Y}=n=50$ (since $Y_{n_{Y}}=Y_{n}$), $\gamma=0.05$, $\alpha_{T}=\alpha$ (only for KFSD$_{tri}$), $n_{Z}=6n$, $\delta=0.05$, $r=\alpha$, desired FAP = 0.10. Moreover, as introduced in Section \ref{sec:kfsd}, for these methods we consider 9 percentiles to set $\sigma$ in KFSD. The way in which we propose to choose the most suitable percentile for outlier detection is presented below. 
\end{compactenum}

\begin{figure}[!htbp]
\centering
\includegraphics[scale=.4]{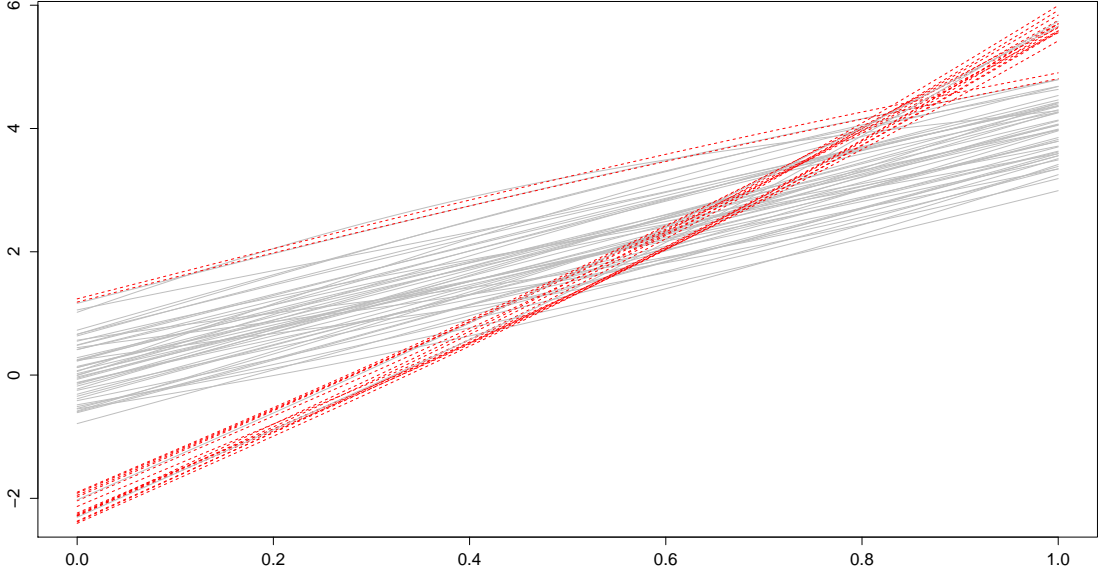}
\captionsetup{width=0.8\textwidth}
\caption{Example of a training sample of peripheral curves for a contaminated data set generated by MM1 with $\alpha=0.05$. The solid and shaded curves are the original curves (both normal and outliers). The dashed curves are the peripheral curves to use as training sample.}
\label{fig:trainPeri}
\end{figure}

In supervised classification, the availability of training curves with known class memberships makes possible the definition of some natural procedures to set $\sigma$ for KFSD, such as cross-validation. However, in an outlier detection problem, it is common to have no information whether curves are normal or outliers. Therefore, training procedures are not immediately available.\\
\indent We propose to overcome this drawback by obtaining a ``training sample of peripheral curves", and then choosing the percentile that ranks better these peripheral curves as final percentile for KFSD in KFSD$_{smo}$, KFSD$_{tri}$ and KFSD$_{wei}$. We now describe this procedure, which is based on $J$ replications. Let $Y_{n}$ be the functional data set on which outlier detection has to be done and let $Y_{(n)} = \left\{y_{(1)}, \ldots, y_{(n)}\right\}$ be the depth-based ordered version of $Y_{n}$, where $y_{(1)}$ and $y_{(n)}$ are the curves with minimum and maximum depth, respectively. The steps to obtain a set of peripheral curves are the following:

\begin{compactenum}[I.]
\item\label{ite:i} Let $\left\{p_{1},\ldots,p_{K}\right\}$ be the set of percentiles in use (in our case, as explained in Section \ref{sec:kfsd}, $p_{k}=(10k)\%$, $k \in \left\{1,\ldots,K=9\right\})$, and choose randomly a percentile from the set. For the $j$th replication, $j \in \left\{1,\ldots,J\right\}$, denote the selected percentile as $p^{j}$. We use $J=20$ in the rest of the paper. 

\item\label{ite:ii} Using $p^{j}$, compute $KFSD_{p^{j}}(y_{i},Y_{n})$, $i=1, \ldots, n$, where the notation $KFSD_{p^{j}}(\cdot,\cdot)$ is used to describe what percentile is used. For the $j$th replication, denote the KFSD-based ordered curves as $y_{(1),j}, \ldots, y_{(n),j}$.

\item\label{ite:iii} Take $y_{(1),j}, \ldots, y_{(l_{j}),j}$, where $l_{j} \sim Bin(n,\frac{1}{n})$. Apply the smoothing step described in Section \ref{sec:odp} to these curves. For the smoothing step, we use $\Sigma_{Y_{n}}$ and $\gamma=0.05$. For the $j$th replication, denote the peripheral and smoothed curves as $y_{(1),j}^{*}, \ldots, y_{(l_{j}),j}^{*}$.

\item\label{ite:iv} Repeat $J$ times steps \ref{ite:i}.-\ref{ite:iii}. to obtain a collection of $L=\sum_{j=1}^{J}l_{j}$ peripheral curves, say $Y_{L}$ (for an example, see Figure \ref{fig:trainPeri}).
\end{compactenum}

Next, $Y_{L}$ acts as training sample according to the following steps: for each $y_{(i),j}^{*} \in Y_{L}$, ($i \leq l_{j}$), and $p_{k} \in \left\{p_{1},\ldots,p_{K}\right\}$, compute $KFSD_{p_{k}}(y_{(i),j}^{*},Y_{-(i),j})$, where $Y_{-(i),j} = Y_{n} \setminus \left\{y_{(i),j}\right\}$. At the end, a $L \times K$ matrix is obtained, say $D_{LK}=\left\{d_{lk}\right\}_{\substack{l=1,\ldots,L\\k=1,\ldots,K}}$, whose $k$th column is composed of the KFSD values of the $L$ training peripheral curves when the $k$th percentile is employed in KFSD. Next, let $r_{lk}$ be the rank of $d_{lk}$ in the vector $\left(KFSD_{p_{k}}(y_{1},Y_{n}),\ldots,KFSD_{p_{k}}(y_{n},Y_{n}),\right.$ $\left. d_{lk} \right)$, e.g., $r_{lk}$ is equal to 1 or $n+1$ if $d_{lk}$ is the minimum or the maximum value in the vector, respectively. Let $R_{LK}$ be the result of this transformation of $D_{LK}$, and sum the elements of each column, obtaining a $K$-dimensional vector, say $\mathbf{R}_{K}$. Since the goal is to assign ranks as low as possible to the peripheral curves, choose the percentile associated to the minimum value of $\mathbf{R}_{K}$. When a tie is observed, we break it randomly.\\
\indent The comparison among methods is performed in terms of both correct and false outlier detection percentages, which are reported in Tables \ref{tab:MM01_I}-\ref{tab:MM06_I}. To ease the reading of the tables, for each model and $\alpha$, we report in bold the 5 best methods in terms of correct outlier detection percentage (c).\footnote{In presence of tie, the method with lower false outlier detection percentage (f) is preferred.} For each model, if a method is among the 5 best ones for both contamination probabilities $\alpha$, we report its label in bold.

\begin{table}[!htbp]
\parbox{.475\textwidth}{
\captionsetup{justification=justified,width=0.475\textwidth}
\caption{MM1, $\alpha=\left\{0.02,0.05\right\}$. Correct (c) and false (f) outlier detection percentages of FBP, B$_{tri}$, B$_{wei}$, FBG, FHD, KFSD$_{smo}$, KFSD$_{tri}$ and KFSD$_{wei}$.}
\centering
\scalebox{.60}{
\begin{tabular}{l|cc|cc}
\hline
& \multicolumn{2}{c}{$\alpha=0.02$} & \multicolumn{2}{|c}{$\alpha=0.05$}\\ 
\hline
& c & f & c & f\\ 
\hline
FBP+FMD  & 44.34 & 1.23 & 43.86 & 0.73\\ 
FBP+HMD  & \textbf{74.53} & 0.94 & 72.81 & 0.61\\ 
FBP+RTD  & 61.32 & 0.57 & 63.16 & 0.31\\ 
FBP+IDD  & 55.66 & 0.61 & 61.84 & 0.34\\ 
FBP+MBD  & 49.06 & 1.33 & 50.44 & 0.69\\ 
FBP+FSD  & 62.26 & 0.67 & 61.84 & 0.40\\ 
FBP+KFSD & 66.04 & 0.86 & \textbf{74.12} & 0.44\\
\hline 
B$_{tri}$+FMD   & 0.00 & 0.98 & 0.00 & 1.82 \\ 
B$_{tri}$+HMD   & 66.98 & 1.45 & 57.89 & 1.47 \\ 
B$_{tri}$+RTD   & 10.38 & 1.78 & 14.91 & 1.76 \\ 
B$_{tri}$+IDD   & 10.38 & 1.55 & 11.84 & 1.74 \\ 
B$_{tri}$+MBD   & 0.00 & 0.51 & 0.00 & 1.49 \\ 
B$_{tri}$+FSD   & 2.83 & 0.76 & 5.26 & 1.17 \\ 
B$_{tri}$+KFSD  & 70.75 & 1.43 & 58.77 & 1.40 \\
\hline
B$_{wei}$+FMD & 0.00 & 1.29 & 0.00 & 1.49 \\ 
B$_{wei}$+HMD & 71.70 & 1.02 & 47.37 & 0.65 \\ 
B$_{wei}$+RTD  & 13.21 & 2.04 & 13.60 & 1.78 \\ 
B$_{wei}$+IDD & 17.92 & 1.82 & 10.53 & 1.55 \\ 
B$_{wei}$+MBD  & 0.00 & 1.08 & 0.00 & 1.40 \\ 
B$_{wei}$+FSD & 2.83 & 1.39 & 3.95 & 1.07 \\ 
B$_{wei}$+KFSD & 61.32 & 0.88 & 55.26 & 0.48 \\        
\hline 
\textbf{FBG} & \textbf{100.00} & 2.27 & \textbf{97.81} & 2.37\\
\hline 
FHD & 48.11 & 1.00 & 73.68 & 2.77\\    
\hline
\textbf{KFSD$_{smo}$} & \textbf{89.62} & 4.50 & \textbf{85.09} & 2.58 \\ 
\textbf{KFSD$_{tri}$} & \textbf{89.62} & 4.92 & \textbf{92.11} & 4.40 \\ 
\textbf{KFSD$_{wei}$} & \textbf{97.17} & 9.44 & \textbf{96.93} & 6.54 \\
\hline
\end{tabular}}
\label{tab:MM01_I}
}
\hfill
\parbox{.475\textwidth}{
\captionsetup{justification=justified,width=0.475\textwidth}
\caption{MM2, $\alpha=\left\{0.02,0.05\right\}$. Correct (c) and false (f) outlier detection percentages of FBP, B$_{tri}$, B$_{wei}$, FBG, FHD, KFSD$_{smo}$, KFSD$_{tri}$ and KFSD$_{wei}$.}
\centering
\scalebox{.60}{
\begin{tabular}{l|cc|cc}
\hline
& \multicolumn{2}{c}{$\alpha=0.02$} & \multicolumn{2}{|c}{$\alpha=0.05$}\\ 
\hline
& c & f & c & f\\ 
\hline
FBP+FMD  & 99.09 & 1.08 & \textbf{96.39} & 0.84\\ 
FBP+HMD  & 96.36 & 0.96 & 96.39 & 0.88\\ 
FBP+RTD  & \textbf{99.09} & 0.61 & 94.78 & 0.25\\ 
FBP+IDD  & 99.09 & 0.70 & 95.18 & 0.38\\ 
FBP+MBD  & 99.09 & 1.06 & \textbf{96.39} & 0.82\\ 
FBP+FSD  & \textbf{99.09} & 0.57 & 94.78 & 0.36\\ 
FBP+KFSD & 98.18 & 0.63 & 93.98 & 0.36\\ 
\hline
B$_{tri}$+FMD & 0.00 & 1.06 & 0.00 & 1.96 \\ 
B$_{tri}$+HMD & 95.45 & 1.51 & \textbf{96.79} & 1.68 \\ 
B$_{tri}$+RTD & 1.82 & 1.92 & 6.83 & 2.61 \\ 
B$_{tri}$+IDD & 5.45 & 1.60 & 7.63 & 1.94 \\ 
B$_{tri}$+MBD & 0.00 & 0.98 & 0.40 & 2.10 \\ 
B$_{tri}$+FSD & 4.55 & 1.06 & 5.22 & 1.62 \\ 
B$_{tri}$+KFSD & 97.27 & 1.60 & 95.18 & 1.52 \\ 
\hline        
B$_{wei}$+FMD & 0.00 & 1.27 & 0.00 & 1.52 \\ 
B$_{wei}$+HMD & 95.45 & 1.02 & 86.35 & 0.36 \\ 
B$_{wei}$+RTD & 5.45 & 2.21 & 8.43 & 2.84 \\ 
B$_{wei}$+IDD & 7.27 & 1.49 & 9.64 & 2.36 \\ 
B$_{wei}$+MBD & 0.00 & 1.27 & 0.40 & 1.49 \\ 
B$_{wei}$+FSD & 8.18 & 1.39 & 4.02 & 1.37 \\ 
B$_{wei}$+KFSD & 95.45 & 0.96 & 79.52 & 0.51 \\ 
\hline
FBG & 8.18 & 3.07 & 4.42 & 2.95\\
\hline 
FHD & 7.27 & 1.88 & 12.45 & 5.66\\ 
\hline
KFSD$_{smo}$ & \textbf{100.00} & 3.91 & 95.18 & 2.76 \\  
\textbf{KFSD$_{tri}$} & \textbf{100.00} & 5.19 & \textbf{97.99} & 4.84 \\
\textbf{KFSD$_{wei}$} & \textbf{100.00} & 9.20 & \textbf{99.60} & 6.48 \\
\hline
\end{tabular}}
\label{tab:MM02_I}
}
\end{table}

\begin{table}[!htbp]
\parbox{.475\textwidth}{
\captionsetup{justification=justified,width=0.475\textwidth}
\caption{MM3, $\alpha=\left\{0.02,0.05\right\}$. Correct (c) and false (f) outlier detection percentages of FBP, B$_{tri}$, B$_{wei}$, FBG, FHD, KFSD$_{smo}$, KFSD$_{tri}$ and KFSD$_{wei}$.}
\centering
\scalebox{.60}{
\begin{tabular}{l|cc|cc}
\hline
& \multicolumn{2}{c}{$\alpha=0.02$} & \multicolumn{2}{|c}{$\alpha=0.05$}\\ 
\hline
& c & f & c & f\\ 
\hline
FBP+FMD  & 65.69 & 0.92 & 49.19 & 0.97\\ 
\textbf{FBP+HMD}  & \textbf{89.22} & 0.57 & \textbf{85.89} & 0.63\\ 
FBP+RTD  & 86.27 & 0.45 & 76.61 & 0.34\\ 
FBP+IDD  & 79.41 & 0.51 & 70.56 & 0.38\\ 
FBP+MBD  & 74.51 & 0.88 & 59.27 & 0.84\\ 
FBP+FSD  & 79.41 & 0.51 & 73.79 & 0.42\\ 
\textbf{FBP+KFSD} & \textbf{89.22} & 0.57 & \textbf{83.06} & 0.59\\
\hline
B$_{tri}$+FMD & 2.94 & 0.73 & 5.24 & 1.22 \\ 
B$_{tri}$+HMD & 57.84 & 1.57 & 53.63 & 1.56 \\ 
B$_{tri}$+RTD & 15.69 & 1.76 & 21.37 & 1.81 \\ 
B$_{tri}$+IDD & 20.59 & 1.65 & 20.56 & 1.70 \\ 
B$_{tri}$+MBD & 0.98 & 1.06 & 3.23 & 1.54 \\ 
B$_{tri}$+FSD & 16.67 & 1.14 & 17.34 & 1.22 \\ 
B$_{tri}$+KFSD & 57.84 & 1.63 & 49.19 & 1.52 \\ 
\hline
B$_{wei}$+FMD & 2.94 & 1.10 & 3.63 & 0.84 \\ 
B$_{wei}$+HMD & 60.78 & 1.25 & 42.74 & 0.76 \\ 
B$_{wei}$+RTD & 15.69 & 1.92 & 17.34 & 1.73 \\ 
B$_{wei}$+IDD & 23.53 & 1.33 & 14.52 & 1.22 \\ 
B$_{wei}$+MBD & 0.98 & 1.29 & 2.82 & 1.14 \\ 
B$_{wei}$+FSD & 15.69 & 1.16 & 12.10 & 0.84 \\ 
B$_{wei}$+KFSD & 56.86 & 1.12 & 41.53 & 0.67 \\ 
\hline
FBG & 86.27 & 2.65 & \textbf{78.63} & 1.73\\
\hline 
FHD & 49.02 & 1.02 & 65.73 & 2.88\\  
\hline
KFSD$_{smo}$ & \textbf{89.22} & 3.90 & 73.79 & 2.95 \\    
\textbf{KFSD$_{tri}$} & \textbf{90.20} & 4.63 & \textbf{83.47} & 4.71 \\
\textbf{KFSD$_{wei}$} & \textbf{97.06} & 8.96 & \textbf{90.32} & 6.50 \\
\hline
\end{tabular}}
\label{tab:MM03_I}
}
\hfill
\parbox{.475\textwidth}{
\captionsetup{justification=justified,width=0.475\textwidth}
\caption{MM4, $\alpha=\left\{0.02,0.05\right\}$. Correct (c) and false (f) outlier detection percentages of FBP, B$_{tri}$, B$_{wei}$, FBG, FHD, KFSD$_{smo}$, KFSD$_{tri}$ and KFSD$_{wei}$.}
\centering
\scalebox{.60}{
\begin{tabular}{l|cc|cc}
\hline
& \multicolumn{2}{c}{$\alpha=0.02$} & \multicolumn{2}{|c}{$\alpha=0.05$}\\ 
\hline
& c & f & c & f\\ 
\hline
FBP+FMD & 1.02 & 0.00 & 0.00 & 0.00\\ 
FBP+HMD & 6.12 & 0.00 & 1.60 & 0.02\\ 
FBP+RTD & 0.00 & 0.00 & 0.00 & 0.00\\ 
FBP+IDD & 0.00 & 0.00 & 0.00 & 0.00\\ 
FBP+MBD & 0.00 & 0.00 & 0.00 & 0.00\\ 
FBP+FSD & 0.00 & 0.00 & 0.00 & 0.00\\ 
FBP+KFSD & 2.04 & 0.00 & 0.80 & 0.00\\
\hline 
B$_{tri}$+FMD & 60.20 & 0.16 & \textbf{47.60} & 0.11 \\ 
B$_{tri}$+HMD & 41.84 & 0.04 & 18.80 & 0.17 \\ 
B$_{tri}$+RTD & 54.08 & 1.16 & 34.80 & 0.82 \\ 
B$_{tri}$+IDD & 55.10 & 1.02 & 37.20 & 0.59 \\ 
\textbf{B$_{tri}$+MBD} & \textbf{64.29} & 0.14 & \textbf{46.40} & 0.13 \\ 
B$_{tri}$+FSD & \textbf{68.37} & 0.14 & 45.60 & 0.08 \\ 
B$_{tri}$+KFSD & 58.16 & 0.20 & 28.00 & 0.13 \\ 
\hline
B$_{wei}$+FMD & 51.02 & 0.12 & 23.60 & 0.00 \\ 
B$_{wei}$+HMD & 38.78 & 0.06 & 10.80 & 0.02 \\ 
B$_{wei}$+RTD & 37.76 & 0.49 & 25.20 & 0.15 \\ 
B$_{wei}$+IDD & 43.88 & 0.67 & 28.00 & 0.42 \\ 
B$_{wei}$+MBD & 56.12 & 0.10 & 25.20 & 0.02 \\ 
B$_{wei}$+FSD & 63.27 & 0.06 & 29.20 & 0.00 \\ 
B$_{wei}$+KFSD & 58.16 & 0.12 & 21.20 & 0.00 \\ 
\hline
FBG & 9.18 & 0.53 & 6.80 & 1.09\\ 
\hline
FHD & 51.02 & 1.02 & 37.60 & 4.34\\ 
\hline
\textbf{KFSD$_{smo}$} & \textbf{87.76} & 2.16 & \textbf{50.00} & 1.24 \\    
\textbf{KFSD$_{tri}$} & \textbf{91.84} & 3.00 & \textbf{64.80} & 2.91 \\
\textbf{KFSD$_{wei}$} & \textbf{95.92} & 5.08 & \textbf{62.00} & 3.35 \\
\hline
\end{tabular}}
\label{tab:MM04_I}
}
\end{table}

\begin{table}[!htbp]
\parbox{.475\textwidth}{
\captionsetup{justification=justified,width=0.475\textwidth}
\caption{MM5, $\alpha=\left\{0.02,0.05\right\}$. Correct (c) and false (f) outlier detection percentages of FBP, B$_{tri}$, B$_{wei}$, FBG, FHD, KFSD$_{smo}$, KFSD$_{tri}$ and KFSD$_{wei}$.}
\centering
\scalebox{.60}{
\begin{tabular}{l|cc|cc}
\hline
& \multicolumn{2}{c}{$\alpha=0.02$} & \multicolumn{2}{|c}{$\alpha=0.05$}\\ 
\hline
& c & f & c & f\\ 
\hline
FBP+FMD  & 55.56 & 0.00 & 54.00 & 0.00\\ 
FBP+HMD  & 66.67 & 0.00 & 68.40 & 0.04\\ 
FBP+RTD  & 57.58 & 0.00 & 54.40 & 0.00\\ 
FBP+IDD  & 52.53 & 0.00 & 56.00 & 0.00\\ 
FBP+MBD  & 55.56 & 0.00 & 55.20 & 0.00\\ 
FBP+FSD  & 55.56 & 0.00 & 55.60 & 0.00\\ 
FBP+KFSD & 60.61 & 0.00 & 59.20 & 0.00\\
\hline 
B$_{tri}$+FMD & 3.03 & 0.18 & 2.80 & 0.44 \\ 
\textbf{B$_{tri}$+HMD} & \textbf{97.98} & 0.12 & \textbf{92.40} & 0.11 \\ 
B$_{tri}$+RTD & 16.16 & 1.06 & 20.00 & 1.03 \\ 
B$_{tri}$+IDD & 18.18 & 1.06 & 16.00 & 1.07 \\ 
B$_{tri}$+MBD & 2.02 & 0.16 & 3.20 & 0.32 \\ 
B$_{tri}$+FSD & 29.29 & 0.18 & 27.20 & 0.23 \\ 
B$_{tri}$+KFSD & 93.94 & 0.24 & \textbf{92.40} & 0.21 \\
\hline 
B$_{wei}$+FMD & 3.03 & 0.29 & 2.40 & 0.23 \\ 
B$_{wei}$+HMD & \textbf{93.94} & 0.08 & 73.60 & 0.00 \\ 
B$_{wei}$+RTD & 15.15 & 1.06 & 17.60 & 1.12 \\ 
B$_{wei}$+IDD & 25.25 & 0.98 & 20.00 & 0.99 \\ 
B$_{wei}$+MBD & 2.02 & 0.20 & 3.60 & 0.21 \\ 
B$_{wei}$+FSD & 29.29 & 0.14 & 21.60 & 0.13 \\ 
B$_{wei}$+KFSD & 83.84 & 0.08 & 72.00 & 0.04 \\
\hline
FBG & 0.00 & 1.02 & 0.40 & 0.04\\
\hline 
FHD & 4.04 & 1.96 & 12.80 & 5.64\\
\hline
\textbf{KFSD$_{smo}$} & \textbf{98.99} & 1.82 & \textbf{94.00} & 0.44 \\      
\textbf{KFSD$_{tri}$} & \textbf{98.99} & 2.61 & \textbf{98.00} & 2.11 \\ 
\textbf{KFSD$_{wei}$} & \textbf{100.00} & 4.61 & \textbf{98.40} & 2.11 \\ 
\hline  
\end{tabular}}
\label{tab:MM05_I}
}
\hfill
\parbox{.475\textwidth}{
\captionsetup{justification=justified,width=0.475\textwidth}
\caption{MM6, $\alpha=\left\{0.02,0.05\right\}$. Correct (c) and false (f) outlier detection percentages of FBP, B$_{tri}$, B$_{wei}$, FBG, FHD, KFSD$_{smo}$, KFSD$_{tri}$ and KFSD$_{wei}$.}
\centering
\scalebox{.60}{
\begin{tabular}{l|cc|cc}
\hline
& \multicolumn{2}{c}{$\alpha=0.02$} & \multicolumn{2}{|c}{$\alpha=0.05$}\\ 
\hline
& c & f & c & f\\ 
\hline
FBP+FMD  & 48.42 & 0.00 & 44.19 & 0.00 \\ 
FBP+HMD  & 60.00 & 0.18 & \textbf{62.92} & 0.00\\ 
FBP+RTD  & 55.79 & 0.00 & 54.68 & 0.00\\ 
FBP+IDD  & 46.32 & 0.00 & 40.07 & 0.00\\ 
FBP+MBD  & 48.42 & 0.00 & 45.69 & 0.00\\ 
FBP+FSD  & 52.63 & 0.00 & 52.43 & 0.00\\ 
FBP+KFSD & 57.89 & 0.00 & 56.93 & 0.00\\
\hline 
B$_{tri}$+FMD & 29.47 & 0.22 & 33.71 & 0.32 \\ 
B$_{tri}$+HMD & \textbf{71.58} & 0.24 & 45.69 & 0.15 \\ 
B$_{tri}$+RTD & 35.79 & 0.82 & 31.09 & 0.51 \\ 
B$_{tri}$+IDD & 38.95 & 0.37 & 35.96 & 0.74 \\ 
B$_{tri}$+MBD & 29.47 & 0.24 & 31.09 & 0.32 \\ 
B$_{tri}$+FSD & 52.63 & 0.20 & 43.82 & 0.19 \\ 
B$_{tri}$+KFSD & \textbf{71.58} & 0.22 & 50.56 & 0.21 \\
\hline 
B$_{wei}$+FMD & 23.16 & 0.24 & 19.48 & 0.08 \\ 
B$_{wei}$+HMD & 68.42 & 0.12 & 35.96 & 0.00 \\ 
B$_{wei}$+RTD & 38.95 & 0.69 & 24.34 & 0.51 \\ 
B$_{wei}$+IDD & 33.68 & 0.59 & 25.09 & 0.40 \\ 
B$_{wei}$+MBD & 24.21 & 0.18 & 19.85 & 0.13 \\ 
B$_{wei}$+FSD & 47.37 & 0.16 & 27.72 & 0.08 \\ 
B$_{wei}$+KFSD & 66.32 & 0.12 & 44.19 & 0.06 \\ 
\hline
FBG & 17.89 & 0.02 & 14.98 & 0.06\\ 
\hline 
FHD & 52.63 & 1.02 & \textbf{61.80} & 2.85\\ 
\hline
\textbf{KFSD$_{smo}$} & \textbf{91.58} & 2.08 & \textbf{71.16} & 0.95 \\    
\textbf{KFSD$_{tri}$} & \textbf{93.68} & 2.69 & \textbf{82.02} & 2.49 \\
\textbf{KFSD$_{wei}$} & \textbf{96.84} & 4.69 & \textbf{83.15} & 2.75 \\
\hline
\end{tabular}}
\label{tab:MM06_I}
}
\end{table}

The results in Tables \ref{tab:MM01_I}-\ref{tab:MM06_I} show that:

\begin{compactenum}
\item KFSD$_{tri}$ and KFSD$_{wei}$ are always among the 5 best methods. KFSD$_{smo}$ is among the 5 best methods 10 times over 12, but when its performance is not among the 5 best, it is neither extremely far from the fifth method (MM2, $\alpha=0.05$: 95.18\% against 96.79\%; MM3, $\alpha=0.05$: 73.79\% against 78.63\%). The rest of the methods are among the 5 best procedures at most 4 times over 12 (FBP+HMD and B$_{tri}$+HMD).

\item Regarding MM5 and MM6, our procedures are clearly the best options in terms of correct detection (c), and in the following order: KFSD$_{wei}$, KFSD$_{tri}$ and KFSD$_{smo}$. In general, this pattern is observed overall the simulation study. Note that for MM6 and $\alpha=0.02$ we observe the best relative performances of KFSD$_{smo}$, KFSD$_{tri}$ and KFSD$_{wei}$, i.e., 91.58\%, 93.68\% and 96.84\%, respectively, against 71.58\% of the fourth best method (B$_{wei}$+KFSD), that is, we observe at least 20\% differences.

\item About MM3, KFSD$_{wei}$ is clearly the best method in terms of correct detection, however at the price of having a greater false detection (f). This is in general the main weak point of KFSD$_{smo}$, KFSD$_{tri}$ and KFSD$_{wei}$. As for correct detection, we observe a overall pattern in our methods in false detection, but in an opposite way, indicating therefore a trade-off between c and f. Relative high false detection percentages are however something expected in KFSD$_{smo}$, KFSD$_{tri}$ and KFSD$_{wei}$ since these methods are based on the definition of a desired false alarm probability, which is equal to 10\% in this study. Concerning MM2, we observe similar results to MM3, but in this case the performances of the best methods in terms of correct detection (KFSD$_{smo}$, KFSD$_{tri}$, KFSD$_{wei}$, FBP-based methods and B$_{tri}$ when used with local depths) are closer to each other.

Finally, there are only 2 cases in which a competitor outperforms all our methods, and it is FBAG under MM1 and both $\alpha$. However, this procedure does not show a behavior as stable as KFSD$_{smo}$, KFSD$_{tri}$ and KFSD$_{wei}$ do. Indeed, FBAG shows poor performances under other models, e.g., MM2.
\end{compactenum}

In summary, the above results and remarks show that the proposed KFSD-based procedures are the best methods in detecting outliers for the considered models. Moreover, KFSD$_{tri}$ seems the most reasonable choice to balance the mentioned trade-off between c and f. In terms of correct detection, KFSD$_{wei}$ slightly outperforms KFSD$_{tri}$, which however shows very good and stable performances when compared with the remaining methods. In terms of false detection, KFSD$_{tri}$ considerably improves on KFSD$_{wei}$, especially under some models (e.g., see MM2).

\begin{figure}[!htbp]
\centering
\includegraphics[scale=.4]{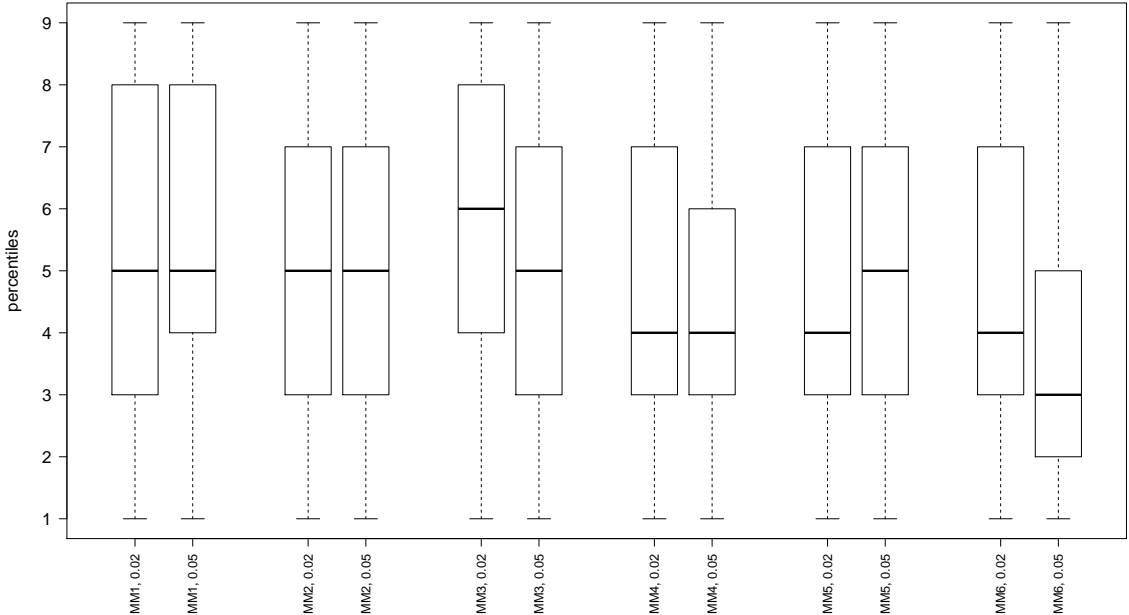}
\captionsetup{width=0.8\textwidth}
\caption{Boxplots of the percentiles selected in the training steps of the simulation study for KFSD$_{smo}$, KFSD$_{tri}$ and KFSD$_{wei}$.}
\label{fig:simBoxPlot}
\end{figure}

In Figure \ref{fig:simBoxPlot} we report a series of boxplots summarizing which percentiles have been selected in the training steps for KFSD$_{smo}$, KFSD$_{tri}$ and KFSD$_{wei}$, and the following general remarks can be made. First, MM6 is the mixture model for which lower percentiles have been selected, and it is also a scenario in which our methods considerably outperform their competitors. The need for a more local approach for MM6-data may explain the two observed facts about this mixture model. Second, lower and more local percentiles have been chosen for mixture models with nonlinear mean functions (MM4, MM5 and MM6) than for mixture models with linear mean functions (MM1, MM2 and MM3). Finally, the percentiles selected by means of the proposed training procedure seem to vary among data sets. However, except for MM3 and $\alpha=0.02$, at least for half of the data sets a percentile not greater than the median has been chosen, which implies at most  a moderately local approach. 

\section{REAL DATA STUDY: NITROGEN OXIDES (NO$_{x}$) DATA}
\label{sec:NOx}
Besides simulated data, we consider a real data set which consists in nitrogen oxides (NO$_{x}$) emission level daily curves measured every hour close to an industrial area in Poblenou (Barcelona) and is available in the \texttt{R} package \texttt{fda.usc} (\citeauthor{FebOvi2012} \citeyear{FebOvi2012}). Outlier detection on this data set was first performed by \cite{FebGalGon2008} where these authors proposed B$_{tri}$ and B$_{wei}$. We carry on their study considering more methods and depths. 

\begin{figure}[!htbp]
\centering
\includegraphics[scale=.4]{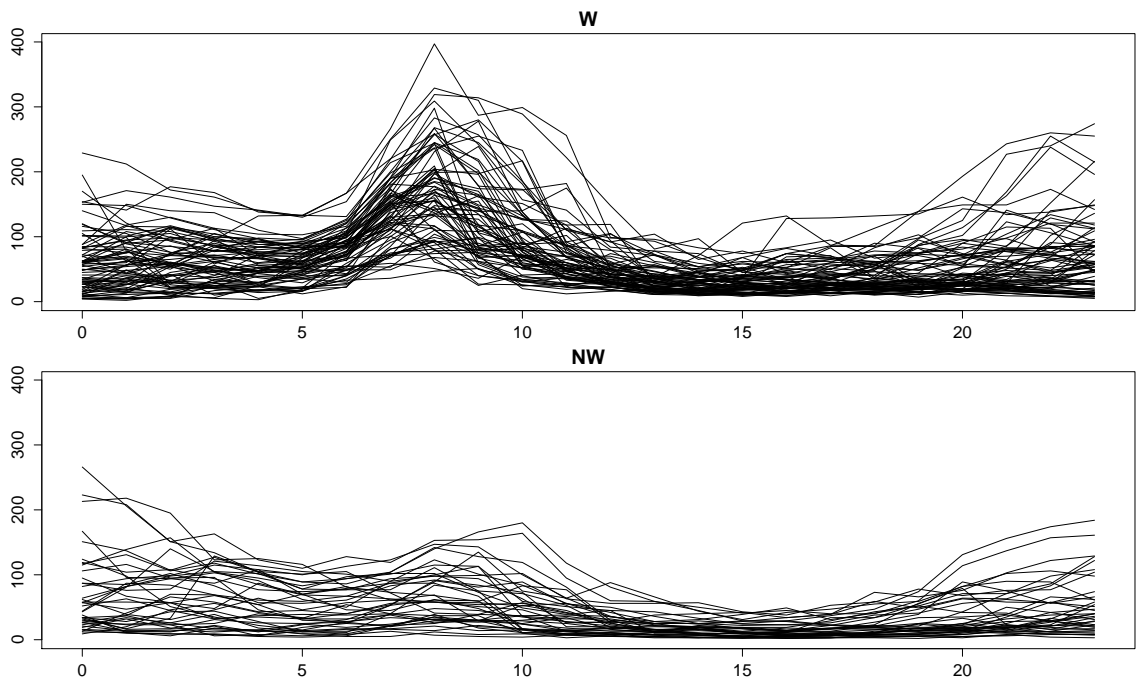}
\captionsetup{width=0.8\textwidth}
\caption{NO$_{x}$ data: working (top) and non working (bottom) day curves.}
\label{fig:NOx}
\end{figure}

NO$_{x}$ are one of the most important pollutants, and it is important to identify outlying trajectories because these curves may compromise any statistical analysis or be of special interest for further analysis and to implement environmental political countermeasures. The NO$_{x}$ levels that we consider were measured in $\mu g/m^{3}$ every hour of every day for the period 23/02/2005-26/06/2005. Only for 115 days of the period are available the 24 measurements, and these are the days that compose the final NO$_{x}$ data set. Moreover, following \cite{FebGalGon2008}, since the NO$_{x}$ data set includes working as well as nonworking days, it seems more appropriate to consider a first sample of 76 working day curves (from now on, W) and a second sample of 39 nonworking day curves (from now on, NW). Both W and NW are showed in Figure \ref{fig:NOx}, where it is possible to appreciate at least two facts that justify the split of the original data set. First, the W curves have in general higher values than NW curves, which can be explained by the greater activity of motor vehicles and industries in a city like Barcelona during working days. Second, both data sets contain curves with peaks, but for W curves the peaks occur roughly around 7-8 a.m. and during many days, whereas for NW curves the peaks occur later and during few days, which again can be explained by the differences between Barcelona's economic activity of working and nonworking days.

\begin{figure}[!htbp]
\centering
\includegraphics[scale=.35]{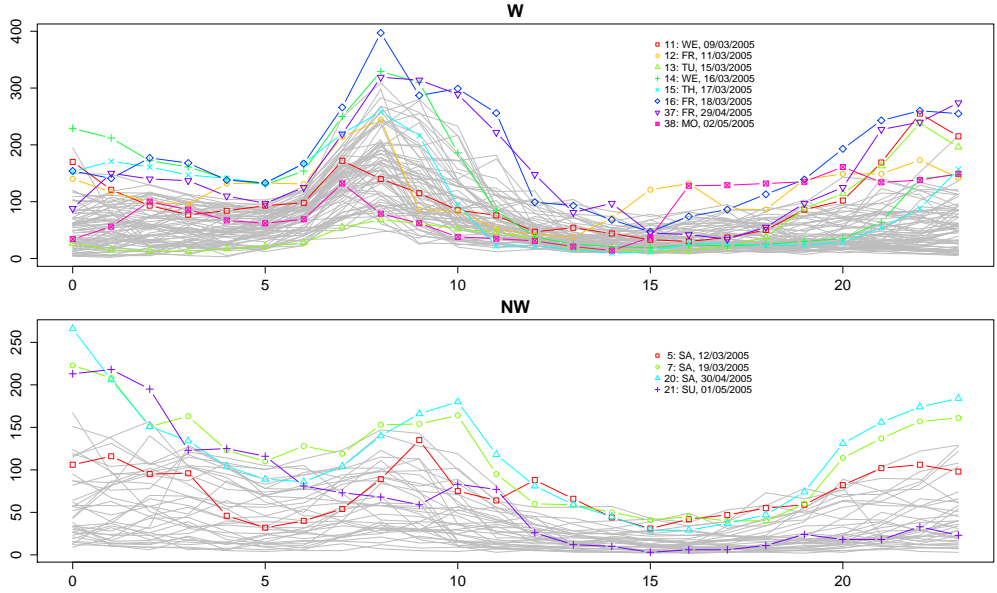}
\captionsetup{width=0.8\textwidth}
\caption{NO$_{x}$ data set, curves detected as outliers in Table \ref{tab:NOx}: working (top) and nonworking (bottom) days.}
\label{fig:NOxIC}
\end{figure}

At first glance, each data set may contain outliers, especially partial outliers in the form of abnormal peaks, and therefore a local depth approach by means of KFSD$_{smo}$, KFSD$_{tri}$ and KFSD$_{wei}$ appears to be a good strategy to detect outliers. Besides them, we do outlier detection with all the methods used in Section \ref{sec:simStudy}. For all the procedures we use the same specifications as in Section \ref{sec:simStudy}, and we assume $\alpha=0.05$. For each method, we report the labels of the curves detected as outliers in Table \ref{tab:NOx} and we highlight these curves in Figure \ref{fig:NOxIC}.

\begin{table}[!htbp]
\captionsetup{justification=justified,width=0.5\textwidth}
\caption{NO$_{x}$ data, Working and Nonworking data sets. Curves detected as outliers by FBP, B$_{tri}$, B$_{wei}$, FBG, FHD, KFSD$_{smo}$, KFSD$_{tri}$ and KFSD$_{wei}$.}
\centering
\scalebox{.64}{
\begin{tabular}{l|c|c}
\hline
& working days & nonworking w days\\ 
\hline
& \multicolumn{2}{c}{detected outliers}\\
\hline
FBP+FMD  & - & -\\ 
FBP+HMD  & 12, 16, 37 & 5, 7, 20, 21\\ 
FBP+RTD  & 37 & 20\\ 
FBP+IDD  & - & 5, 7, 20\\ 
FBP+MBD  & - & -\\ 
FBP+FSD  & 37 & -\\ 
FBP+KFSD & 12, 16, 37 & 5, 7, 20, 21\\
\hline
B$_{tri}$+FMD  & 16, 37 & 7\\          
B$_{tri}$+HMD  & 14, 16, 37 & 7, 20\\     
B$_{tri}$+RTD  & 16 & 7, 20\\          
B$_{tri}$+IDD  & 16, 37 & 7, 20\\  
B$_{tri}$+MBD  & 16, 37 & 7\\          
B$_{tri}$+FSD  & 14, 16, 37 & -\\      
B$_{tri}$+KFSD & 12, 14, 16, 37 & 7, 20\\      
\hline                                                         
B$_{wei}$+FMD  & 16 & 7, 20\\                  
B$_{wei}$+HMD  & 16, 37 & 7, 20\\          
B$_{wei}$+RTD  & 16 & -\\          
B$_{wei}$+IDD  & 16, 37 & 20\\ 
B$_{wei}$+MBD  & 16 & 7\\                  
B$_{wei}$+FSD  & 16, 37 & -\\         
B$_{wei}$+KFSD & 16, 37 & 7, 20\\       
\hline
FBG & 16, 37 & -\\
\hline                                                         
FHD & 12, 14, 16, 37 & 7, 20\\ 
\hline
KFSD$_{smo}$ & 14, 16, 37 & 7, 20, 21\\
KFSD$_{tri}$ & 12, 14, 16, 37 & 7, 20, 21\\
KFSD$_{wei}$ & 11, 12, 13, 14, 15, 16, 37, 38 & 7, 20, 21\\
\hline
\end{tabular}}
\label{tab:NOx}
\end{table}

Concerning W, most of the methods detect as outlier day 37, the Friday at the beginning of the long weekend due to Labor's day in 2005 and whose curve shows a partial outlying behavior before noon and at the end of the day. Another day detected as outlier by many methods is day 16, another Friday before a long weekend, Easter holidays in 2005, and whose curve has the highest morning peak. In addition to curves 16 and 37, KFSD$_{smo}$ detects as outlier curve 14, as other nine methods do, recognizing a seemingly outlying pattern in early hours of the day. Additionally, KFSD$_{tri}$ includes among the outliers also day 12, which may be atypical because of its behavior in early afternoon. Note that both day 12 and 14 are in the week before the above-mentioned Easter holidays. Finally, KFSD$_{wei}$ detects as outliers the greatest number of curves. This last result may appear exaggerated, but all the curves that are outliers according to KFSD$_{wei}$ seem to have some partial deviations from the majority of curves. For example, day 13, whose curve is considered normal by the rest of the procedures, shows a peak at end of the day. Similar peaks can be observed also in other curves detected as outliers by other methods (e.g., days 16 and 37), which means that it may be occurring a masking effect to day 13's detriment, and only KFSD$_{wei}$ points out this possibly outlying feature of the curve. Regarding the training step for KFSD to set $\sigma$, it gives as result the 70\% percentile. Observing the first graph of Figure \ref{fig:NOx}, it can be noticed that some curves have a likely outlying behavior, and this may be the reason why a weakly local approach for KFSD may be adequate enough.

In the case of NW, some methods detect no curves as outliers (e.g., all the FSD-based methods), exclusively three FBP-based methods flag day 5 as outlier, whereas days 7, 20 and 21 are detected as outliers by our methods as well as others. Note that day 7 is the Saturday before Easter and days 20 and 21 are Labor's day eve and the same Labor's day. Days 7 and 20, which have two peaks, at the beginning and end of the day, are also flagged by other twelve and eight methods, respectively, while day 21, which shows a single peak in the first hours of the day, is considered atypical by only two other methods, which happen to be local (FBP+HMD and FBP+KFSD). This last result may be connected with what has been observed at the KFSD training step for selecting the percentile, i.e., the selection of the 30\% percentile. Therefore, KFSD$_{smo}$, KFSD$_{tri}$ and KFSD$_{wei}$ work with a strongly local percentile, and their results partially resemble the ones of the previously mentioned local techniques.

\section{CONCLUSIONS}
\label{sec:conc}
This paper proposes to tackle outlier detection in functional samples using the kernelized functional spatial depth as a tool. In Theorem \ref{th:inPaper01} we presented a probabilistic result allowing to set a KFSD-threshold to identify outliers, but in practice it is necessary to observe two samples to apply Theorem \ref{th:inPaper01}. To overcome this practical limitation, we proposed KFSD$_{smo}$, KFSD$_{tri}$ and KFSD$_{wei}$ which are methods that can be applied when a unique functional sample is available and are based on both a probabilistic approach and smoothed resampling techniques.

We also proposed a new procedure to set the bandwidth $\sigma$ of KFSD that is based on obtaining training samples by means of smoothed resampling techniques. The general idea behind this procedure can be applied to other functional depths or methods with parameters that need to be set.

We investigated the performances of KFSD$_{smo}$, KFSD$_{tri}$ and KFSD$_{wei}$ by means of a simulation study. We focused on challenging scenarios with low magnitude, shape and partial outliers instead of high magnitude outliers. The results support our proposals. Along the simulation study, KFSD$_{smo}$, KFSD$_{tri}$ and KFSD$_{wei}$ attained the largest correct detection performances in most of the analyzed setups, but in some cases they paid a price in terms of false detection. However, KFSD$_{smo}$, KFSD$_{tri}$ and KFSD$_{wei}$ work with a given desired false alarm probability, and therefore higher false detection percentages than their competitors are due to the inherent structure of the methods. We also observed a trade-off between c and f for KFSD$_{smo}$, KFSD$_{tri}$ and KFSD$_{wei}$, and a clear pattern. For these reasons in our opinion KFSD$_{tri}$ should be preferred to KFSD$_{smo}$ or KFSD$_{wei}$ since it performs extremely well in terms of correct detection, while it has lower false detection percentages than KFSD$_{wei}$. Concerning the remaining methods, there are competitors that in few scenarios outperformed our methods. However, in these few cases the differences are not great, and in addition these competitors are not stable across the considered scenarios. 

Furthermore, we also showed that our procedures can be applied in environmental contexts with an example where the goal was to detect outlying NO$_{x}$ curves to identify days possibly characterized by abnormal pollution levels.

To conclude, we present two possible future research lines. First, since KFSD is a depth whose local approach is in part based on the choice of the kernel function, it would be interesting to explore how the choice of different kernels affects the behavior of KFSD. Moreover, each kernel will depend on a bandwidth and a norm. For the selection of the bandwidth, we used a criterion based on the study of the empirical distribution of the sample distances, but alternatives should be investigated, for example an adaptation of the so-called Silverman's rule (\citeauthor{Sil1986} \citeyear{Sil1986}) for selecting the bandwidth of a kernel-based functional depth such as KFSD. For the choice of the norm, a sensitivity study would help in understanding how important is the functional space assumption. Second, since outlier detection can be seen as a special case of cluster analysis (it is a cluster problem with maximum two clusters, and one of them with size much smaller than the other,even 0), a natural step ahead in our research may be the definition of KFSD-based cluster analysis procedures.

%\if1\blind
{
\section*{ACKNOWLEDGMENTS}
The authors would like to thank the editor in chief, the associate editor and an anonymous referee for their helpful comments. This research was partially supported by Spanish Ministry of Science and Innovation grant ECO2011-25706 and by Spanish Ministry of Economy and Competition grant ECO2012-38442.
}
% \fi

\appendix
\section{Appendix}

\subsection{From $FSD(x, Y_{n})$ to $KFSD(x, Y_{n})$}
\label{sec:app01}
To show how to pass from $FSD(x, Y_{n})$ in \eqref{eq:sampleFSD} to $KFSD(x, Y_{n})$ in \eqref{eq:sampleKFSD}, we first show that $FSD(x, Y_{n})$ can be expressed in terms of inner products. We present this result for $n=2$. The norm in \eqref{eq:sampleFSD} can be written as

\begin{equation*}
\begin{array}{ll}
\left\|\sum_{i}^{2}\frac{x-y_{i}}{\|x-y_{i}\|}\right\|^{2} &= \left\|\frac{x-y_{1}}{\|x-y_{1}\|}+\frac{x-y_{2}}{\|x-y_{2}\|}\right\|^{2}\\
&= \left\|\frac{x-y_{1}}{\sqrt{\langle x,x\rangle+\langle y_{1},y_{1}\rangle-2\langle x,y_{1}\rangle}}+\frac{x-y_{2}}{\sqrt{\langle x,x\rangle+\langle y_{2},y_{2}\rangle-2\langle x,y_{2}\rangle}}\right\|^{2}
\end{array}
\end{equation*}

\noindent Let $\delta_{1}=\sqrt{\langle x,x\rangle+\langle y_{1},y_{1}\rangle-2\langle x,y_{1}\rangle}$ and $\delta_{2}=\sqrt{\langle x,x\rangle+\langle y_{2},y_{2}\rangle-2\langle x,y_{2}\rangle}$. Then,

\begin{equation*}
\begin{array}{ll}
\left\|\sum_{i}^{2}\frac{x-y_{i}}{\|x-y_{i}\|}\right\|^{2} &= \left\|\frac{x-y_{1}}{\delta_{1}}+\frac{x-y_{2}}{\delta_{2}}\right\|^{2} \\
&= \left\|\frac{x-y_{1}}{\delta_{1}}\right\|+\left\|\frac{x-y_{2}}{\delta_{2}}\right\| + \frac{2}{\delta_{1}\delta_{2}}\langle x-y_{1},x-y_{2}\rangle \\
&= 2 + \frac{2}{\delta_{1}\delta_{2}}(\langle x,x\rangle+\langle y_{1},y_{2}\rangle-\langle x,y_{1}\rangle-\langle x,y_{2}\rangle)\\
&= \sum_{i,j=1}^{2}\frac{\langle x,x\rangle+\langle y_{i},y_{j}\rangle-\langle x,y_{i}\rangle-\langle x,y_{j}\rangle}{\delta_{i}\delta_{j}},
\end{array}
\end{equation*}

\noindent and apply the embedding map $\phi$ to all the observations of the last expression. According to \eqref{eq:kappa_phi}, this is equivalent to substitute the inner product function with a positive definite and stationary kernel function $\kappa$, which explains the definition of $KFSD(x, Y_{n})$ in \eqref{eq:sampleKFSD} for $n=2$. The generalization of this result to $n>2$ is straightforward.
 
\subsection{Proof of Theorem \ref{th:inPaper01}}
\label{sec:app02}
As explained in Section \ref{sec:odp}, Theorem \ref{th:inPaper01} is a functional extension of a result derived by \cite{CheDanPenBar2009} for KSD, and since they are closely related, next we report a sketch of the proof of Theorem \ref{th:inPaper01}. The proof for KSD is mostly based on an inequality known as \citeauthor{Mcd1989}'s inequality (\citeauthor{Mcd1989} \citeyear{Mcd1989}), which also applies to general probability spaces, and therefore to functional Hilbert spaces. We report this inequality in the next lemma:

\begin{lemma}\label{le:01}
\textbf{(\citeauthor{Mcd1989} \citeyear{Mcd1989} [1.2])} Let $\Omega_{1}, \ldots, \Omega_{n}$ be probability spaces. Let $\mathbf{\Omega} = \prod_{j=1}^{n} \Omega_{j}$ and let $X: \mathbf{\Omega} \rightarrow \mathbb{R}$ be a random variable. For any $j \in \left\{1, \ldots, n\right\}$, let $(\omega_{1}, \ldots, \omega_{j}, \ldots,$ $\omega_{n})$ and $\left(\omega_{1}, \ldots, \hat{\omega}_{j}, \ldots, \omega_{n}\right)$ be two elements of $\mathbf{\Omega}$ that differ only in their $j$th coordinates. Assume that $X$ is uniformly difference-bounded by $\{c_j\}$, that is, for any $j \in \left\{1, \ldots, n\right\}$,

\begin{equation}
\label{eq:lem01_c_j}
\left|X\left(\omega_{1}, \ldots, \omega_{j}, \ldots, \omega_{n}\right)-X\left(\omega_{1}, \ldots, \hat{\omega}_{j}, \ldots, \omega_{n}\right)\right| \leq c_{j}.
\end{equation}

\noindent Then, if $\mathbb{E}[X]$ exists, for any $\tau > 0$

\begin{equation*}
\mathrm{Pr}\left(X-\mathbb{E}[X] \geq \tau \right) \leq \exp \left(\frac{-2\tau^2}{\sum_{j=1}^{n} c_{j}^{2}}\right).
\end{equation*}

\end{lemma}

In order to apply Lemma \ref{le:01} to our problem, define  

\begin{equation}
\label{eq:lem01_X_omega}
X(z_{1},\ldots,z_{n_{Z}}) = - \frac{1}{n_{Z}}\sum_{i=1}^{n_{Z}}g(z_{i},Y_{n_{Y}}|Y_{n_{Y}}),
\end{equation}

\noindent whose expected value is given by

\begin{equation}
\label{eq:lem01_EX}
\mathbb{E}[X] = \mathbb{E}_{z_{i}|Y_{n_{Y}}}\left[- \frac{1}{n_{Z}}\sum_{i=1}^{n_{Z}}g(z_{i},Y_{n_{Y}}|Y_{n_{Y}})\right] = - \mathbb{E}_{z_{1}|Y_{n_{Y}}}\left[g(z_{1},Y_{n_{Y}}|Y_{n_{Y}})\right].
\end{equation}

Now, for any $j \in \left\{1, \ldots, n_{Z}\right\}$ and $\hat{z}_{j} \in \mathbb{H}$, the following inequality holds

\begin{equation*}
\left|X(z_{1},\ldots,z_{j},\ldots,z_{n_{Z}}) - X(z_{1},\ldots,\hat{z}_{j},\ldots,z_{n_{Z}})\right| \leq \frac{1}{n_{Z}},
\end{equation*}

\noindent and it provides assumption \eqref{eq:lem01_c_j} of Lemma \ref{le:01}. Therefore, for any $\tau > 0$

\begin{equation*}
\mathrm{Pr}\left(\mathbb{E}_{z_{1}|Y_{n_{Y}}}\left[g(z_{1},Y_{n_{Y}}|Y_{n_{Y}})\right] - \frac{1}{n_{Z}}\sum_{i=1}^{n_{Z}}g(z_{i},Y_{n_{Y}}|Y_{n_{Y}}) \geq \tau\right) \leq \exp\left(-2n_{Z}\tau^{2}\right),
\end{equation*}

\noindent and by the law of total probability 

\begin{equation*}
\begin{array}{l}
\mathbb{E}\left[\mathrm{Pr}\left(\mathbb{E}_{z_{1}|Y_{n_{Y}}}\left[g(z_{1},Y_{n_{Y}}|Y_{n_{Y}})\right] - \frac{1}{n_{Z}}\sum_{i=1}^{n_{Z}}g(z_{i},Y_{n_{Y}}|Y_{n_{Y}}) \geq \tau\right)\right] \\
= \mathrm{Pr}\left(\mathbb{E}_{z_{1}|Y_{n_{Y}}}\left[g(z_{1},Y_{n_{Y}})\right] - \frac{1}{n_{Z}}\sum_{i=1}^{n_{Z}}g(z_{i},Y_{n_{Y}}) \geq \tau\right) \leq \exp\left(-2n_{Z}\tau^{2}\right)\\
\end{array}
\end{equation*}

Next, setting $\delta = \exp\left(-2n_{Z}\tau^{2}\right)$, and solving for $\tau$, the following result is obtained:

\begin{equation*}
\tau = \sqrt{\frac{\ln 1/\delta}{2n_{Z}}}.
\end{equation*}

\noindent Therefore,

\begin{equation}
\label{eq:pr_z1}
\mathrm{Pr}\left(\mathbb{E}_{z_{1}|Y_{n_{Y}}}\left[g(z_{1},Y_{n_{Y}})\right] \leq \frac{1}{n_{Z}}\sum_{i=1}^{n_{Z}}g(z_{i},Y_{n_{Y}}) + \sqrt{\frac{\ln 1/\delta}{2n_{Z}}}\right) \geq 1-\delta.
\end{equation}

However, Theorem \ref{th:inPaper01} provides a probabilistic upper bound for $\mathbb{E}_{x|Y_{n_{Y}}}\left[g(x,Y_{n_{Y}})\right]$. First, recall that $z_{1} \sim Y_{mix}$ and note that

\begin{equation*}
\mathbb{E}_{(z_{1} \sim Y_{mix})|Y_{n_{Y}}}\left[g\left(z_{1},Y_{n_{Y}}\right)\right] = (1-\alpha)\mathbb{E}_{(z_{1} \sim Y_{nor})|Y_{n_{Y}}}\left[g\left(z_{1},Y_{n_{Y}}\right)\right] + \alpha\mathbb{E}_{(z_{1} \sim Y_{out})|Y_{n_{Y}}}\left[g\left(z_{1},Y_{n_{Y}}\right)\right].
\end{equation*}

\noindent Then, since $\mathbb{E}_{(z_{1} \sim Y_{nor})|Y_{n_{Y}}}\left[g\left(z_{1},Y_{n_{Y}}\right)\right] = \mathbb{E}_{x|Y_{n_{Y}}}\left[g\left(x,Y_{n_{Y}}\right)\right]$, for $\alpha>0$,

\begin{equation}
\label{eq:nor_mix_mix}
\mathbb{E}_{x|Y_{n_{Y}}}\left[g\left(x,Y_{n_{Y}}\right)\right] \leq \frac{1}{1-\alpha}\mathbb{E}_{(z_{1} \sim Y_{mix})|Y_{n_{Y}}}\left[g\left(z_{1},Y_{n_{Y}}\right)\right].
\end{equation}

\noindent Consequently, combining \eqref{eq:pr_z1} and \eqref{eq:nor_mix_mix}, and for $r \geq \alpha$, we obtain

\begin{equation*}
\mathrm{Pr}\left(\mathbb{E}_{x|Y_{n_{Y}}}\left[g(x,Y_{n_{Y}})\right] \leq \frac{1}{1-r}\left[\frac{1}{n_{Z}}\sum_{i=1}^{n_{Z}}g(z_{i},Y_{n_{Y}}) + \sqrt{\frac{\ln 1/\delta}{2n_{Z}}}\right]\right) \geq 1-\delta,
\end{equation*}

\noindent which completes the proof.

\begin{flushright}
$\blacksquare$
\end{flushright}

\bibliographystyle{spbasic}      
\bibliography{paperSERRAReferences}

\end{document}